\documentclass[aps,prd,reprint,nofootinbib,floatfix,longbibliography]{revtex4-2}
\usepackage[table]{xcolor}
\usepackage{graphicx}  
\usepackage{dcolumn}   
\usepackage{bm}        
\usepackage{amssymb}   
\usepackage{blindtext}
\usepackage{float}
\usepackage{amsmath}  
\usepackage [english]{babel}
\usepackage{braket}
\usepackage[utf8]{inputenc}
\usepackage[pdfencoding=auto,psdextra]{hyperref}
\usepackage{url}
\usepackage{hyperref}

\usepackage{orcidlink}
\newcommand{\orcid}[1]{\orcidlink{#1}}

\usepackage{rotating}

\usepackage{cancel}
\usepackage{makecell}

\begin{document}

\title{Nonrelativistic nuclear reduction for tensor couplings \\ in dark matter direct detection and $\mu \to e$ conversion}

\author{Ayala Glick-Magid \orcid{0000-0003-4499-8303}}
\email[E-mail:~]{glickm@uw.edu}
\affiliation{Department of Physics, University of Washington, Seattle, Washington 98195, USA}
\affiliation{Institute for Nuclear Theory, University of Washington, Seattle, Washington 98195, USA}

\begin{abstract}
The nonrelativistic effective field theory (NRET) is widely used in dark matter direct detection and charged-lepton flavor violation studies through $\mu \to e$ conversion. However, existing literature has not fully considered tensor couplings. This study fills this gap by utilizing an innovative tensor decomposition method, extending NRET to incorporate previously overlooked tensor interactions. This development is expected to have a significant impact on ongoing experiments seeking physics beyond the Standard Model and on our understanding of the new-physics interactions. Notably, we identify additional operators in $\mu \to e$ conversion that are absent in scalar and vector couplings. To support further research and experimental analyses, comprehensive tables featuring tensor matrix elements and their corresponding operators are provided.
\end{abstract}

\maketitle

Cosmological and astrophysical observations have established that about 25\% of the energy density in our universe is attributed to dark matter (DM), a form of matter that rarely interacts with regular matter and has not yet been directly observed~\cite{gaitskell2004direct, bertone2005particle, klos2013large}. This matter serves as an explanation for certain gravitational phenomena, such as the angular velocity of gas clouds around galaxies, and the motion of galaxies within clusters, which appear to be operating under the influence of additional mass that cannot be detected.
As DM candidates are naturally predicted by extensions of the Standard Model (SM) of particle physics~\cite{feng2010dark}, unraveling the nature of DM remains a paramount challenge in particle physics and astrophysics.

A prominent DM candidate is weakly-interacting massive particles (WIMPs)~\cite{bertone2005particle}. 
WIMPs interact with quarks, enabling their detection through elastic scattering off nuclei \cite{baudis2012direct}. The momentum transfer in such interactions typically lies around $q \sim 100$ MeV~\cite{menendez2012spin}.
In recent years, direct detection experiments have been introduced that are tailored to this energy range~\cite{hoferichter2015chiral}. Earth-based detectors aim to capture interactions of extraterrestrial DM by measuring the recoil energy of nuclei, indicative of DM scattering events.

Assessing how nuclei respond to WIMPs requires matching WIMP-quark couplings to WIMP-nucleon currents. Given the non-perturbative nature of QCD at low energies, this is best accomplished using effective theories. In the literature, two methods have been utilized: nonrelativistic effective field theory (NRET, also known as NREFT)~\cite{fitzpatrick2013effective, anand2014weakly, gazda2017ab}, and chiral effective field theory ($\chi$EFT)~\cite{Cirigliano2012wimp, menendez2012spin, klos2013large, hoferichter2015chiral}.

In both theories, the coupling structure is solely determined by symmetry considerations. However, early works~\cite{fitzpatrick2013effective, anand2014weakly, hoferichter2015chiral, hoferichter2016analysis} overlooked the tensor symmetry coupling, crucial for understanding the nature of DM-nucleon interactions once detected, analogous to the identification of the $V-A$ structure of weak interactions in the SM during the previous century.
Only recently, the tensor symmetry coupling has been incorporated into $\chi$EFT~\cite{hoferichter2019nuclear}, but it is still missing in NRET.
Due to the extensive use of NRET in various DM search applications, such as software programs (e.g.,~\cite{anand2014weakly, bradley2021software}), numerical calculations (e.g.,~\cite{gazda2017ab, Khaleq2023impact}), direct detection experiments (e.g.,~\cite{Edwards2019digging, PICO2023search}), and forecasts for new measurement opportunities (e.g.,~\cite{Cappiello2023dark}), this gap cannot be ignored.

The utility of NRET extends beyond DM interactions with nuclei, encompassing explorations into other physics beyond the Standard Model (BSM). 
The 2015 Nobel Prize for establishing the flavor oscillations of neutrinos further motivated the search for additional flavor-violation among charged leptons.
A notable charged-lepton flavor-violating (CLFV) process is muon-to-electron ($\mu \to e$) conversion, where a muon transforms into an electron.

Several experiments aim to detect this conversion, such as Mu2e at Fermilab~\cite{bartoszek2015mu2e, abusalma2018expression}, and COMET~\cite{angelique2018comet, comet2020comet} and DeeMe at J-PARC, poised to significantly advance CLFV limits on the branching ratios by four orders of magnitude~\cite{haxton2023nuclear}. These experiments seek evidence of $\mu \to e$ conversion within a nucleus, requiring accurate predictions incorporating all known symmetries. Recent studies underscored the effectiveness of NRET in calculating $\mu \to e$ conversion~\cite{Rule2023nuclear, haxton2023nuclear}, introducing a nuclear-level effective theory for this process, and highlighting the potential of the missing tensor mediators to introduce operators not found in other symmetries.

We have recently developed a method for decomposing fermionic tensor-type interactions based on their antisymmetric nature~\cite{glick2023multipole}. This method has proven to be highly useful in describing the impact of BSM tensor symmetry on different interactions, including precision studies of $\beta$-decay~\cite{glick2023multipole, glick2022formalism}, neutrino scattering~\cite{hoferichter2020coherent}, and more~\cite{glick2023multipole, hoferichter2023improved}, and led to new experiments~\cite{glick2017beta, mardor2018soreq, ohayon2018weak} and tools~\cite{shuai2022determination}. As indicated in~\cite{hoferichter2023improved}, this tensor-decomposing approach can readily provide the missing tensor terms crucial for understanding the nature of DM and CLFV interactions upon their detection.

This letter fills the gap in these two seemingly disparate but framework-sharing BSM searches, employing the tensor-decomposing approach~\cite{glick2023multipole}. We introduce NRET and its tensor completion for DM, followed by $\mu \to e$ conversion.

\paragraph*{DM direct detection.\label{sec: DM}}
DM detection necessitates structure factors for elastic WIMP-nucleus scattering, which are particularly sensitive to the nuclear structure inherent in spin-dependent WIMP-nucleon interactions. Let us consider a contact interaction involving a spin-half WIMP denoted as $\chi$ and a nucleon represented by $N$. In the framework of NRET, the comprehensive Lagrangian density is expressed as follows~\cite{fitzpatrick2013effective}:
\begin{align}
\mathcal{\hat{L}}_{\text{int}} & =\bar{\chi}\mathcal{O}_{\chi}\chi\bar{N}\mathcal{O}_{N}N\text{.}
\end{align}
Here, the properties of the WIMP operator $\mathcal{O}_{\chi}$ and the nucleon operator $\mathcal{O}_{N}$ are determined by enforcing their corresponding symmetries, which may take the form of any of the bilinear covariants: scalar (including pseudoscalar), vector (including axial-vector), and tensor.
By considering the scalar and vector symmetries, it has been demonstrated that the contact Lagrangian density, at leading order in $p/m_{\chi}$ and $k/m_{N}$ (where $p^{\mu}$ and $k^{\mu}$ are the four-momenta of the DM and nucleon, respectively, and $m_{\chi}$ and $m_N$ are their masses), can be expressed in terms of 16 nonrelativistic operators \cite{fitzpatrick2013effective, anand2014weakly}:
\begin{align}
\mathcal{\hat{L}}_{\text{int}} & =\Sigma_{i=1}^{16}c_{i}\mathcal{O}_{i}\bar{\chi}\chi\bar{N}N\text{,}
\end{align}
where the operators $\mathcal{O}_{i}$ are constructed from the DM (nucleon) identity operators $1_{\chi}$ ($1_N$), and the three-vectors:
$i\frac{\vec{q}}{m_{N}}$, $\vec{v}^{\bot}\equiv\frac{\vec{P}}{2m_{\chi}}-\frac{\vec{K}}{2m_{N}}$,
$\vec{S}_{\chi} \equiv \frac{\vec{\sigma}_{\chi}}{2}$ and $\vec{S}_{N}\equiv \frac{\vec{\sigma}_{N}}{2}$.
In these expressions, $q^{\mu}\equiv p_f^{\mu}-p_i^{\mu}=k_i^{\mu}-k_f^{\mu}$ denotes the 4-momentum transfer, $P^{\mu}\equiv p_i^{\mu}+p_f^{\mu}$, $K^{\mu}\equiv k_i^{\mu}+k_f^{\mu}$, $k_i^{\mu}$ ($k_f^{\mu}$) and $p_i^{\mu}$ ($p_f^{\mu}$) are the incoming (outgoing) 4-momenta, and $\vec{\sigma}_{\chi}$ ($\vec{\sigma}_{N}$) is the spin operator of the DM (nucleon).
%
\begin{table*}
\centering
\rotatebox{90}{
\begin{minipage}{\textheight}
  \caption{\label{tab: WIMP Nonrelativistic Reduction}
The second column showcases the tensor Lagrangian densities $\mathcal{L}_{\text{int}}^{j}$ corresponding to WIMP-nucleus scattering, with the index $j$ enumerated in the first column. The terms encompass scalar and vector contributions outlined in \cite{anand2014weakly} ($j \in \left\{1, 2, ..., 20\right\}$), augmented here by the inclusion of previously unaccounted tensor terms. In the third column, the operators resulting from the nonrelativistic reduction between Pauli spinors are presented, while the fourth column illustrates the associated effective interactions expressed in terms of the NRET operators defined in Eq. \eqref{eq: NRET_operators_WIMP}. Further details can be found in \cite{anand2014weakly}.}
\begin{ruledtabular}
\begin{tabular}{cccc}
$j$ & $\mathcal{L}_{\text{int}}^{j}$ & Pauli operator reduction & $\Sigma_{i}c_{i}\mathcal{O}_{i}$
\\
\hline 
21 & $\bar{\chi}\sigma^{\mu\nu}\chi\bar{N}\sigma_{\mu\nu}N$ & $8\frac{\vec{\sigma}_{\chi}}{2}\cdot\frac{\vec{\sigma}_{N}}{2}+\mathit{O}\left(\frac{1}{m^{2}}\right)$ & $8\mathcal{O}_{4}$
\\
22 & $\bar{\chi}\sigma^{\mu\nu}\chi\bar{N}\left(\frac{q_{\mu}}{m_{M}}\gamma_{\nu}-\frac{q_{\nu}}{m_{M}}\gamma_{\mu}\right)N$ 
& $-\frac{iq^{2}}{m_{\chi}m_{M}}1_{\chi}1_{N}
-\frac{4}{m_{M}}\frac{\vec{\sigma}_{\chi}}{2}\cdot\left(\vec{q}\times\vec{v}^{\bot}\right)$ 
& $-i\frac{q^{2}}{m_{M}m_{\chi}}\mathcal{O}_{1}+4i\frac{m_{N}}{m_{M}}\mathcal{O}_{5}$
\\
& & $+\frac{4i}{m_{N}m_{M}}\left[\left(\frac{\vec{\sigma}_{N}}{2}\cdot\vec{q}\right)\left(\frac{\vec{\sigma}_{\chi}}{2}\cdot\vec{q}\right)
-q^{2}\left(\frac{\vec{\sigma}_{\chi}}{2}\cdot\frac{\vec{\sigma}_{N}}{2}\right)
\right]+\mathit{O}\left(\frac{1}{m^{3}}\right)$ 
& $+4i\frac{m_{N}}{m_{M}}\mathcal{O}_{6}-4i\frac{q^{2}}{m_{M}m_{N}}\mathcal{O}_{4}$
\\
23 & $\bar{\chi}\sigma^{\mu\nu}\chi\bar{N}\left(\frac{q_{\mu}}{m_{M}}\frac{K_{\nu}}{m_{M}}-\frac{q_{\nu}}{m_{M}}\frac{K_{\mu}}{m_{M}}\right)N$ & $-2i\frac{m_{N}}{m_{\chi}}\frac{q^{2}}{m_{M}^{2}}1_{\chi}1_{N}-8\frac{m_{N}}{m_{M}^{2}}\frac{\vec{\sigma}_{\chi}}{2}\cdot\left(\vec{q}\times\vec{v}^{\bot}\right)+\mathit{O}\left(\frac{1}{m^{4}}\right)$ & $-2i\frac{m_{N}}{m_{\chi}}\frac{q^{2}}{m_{M}^{2}}\mathcal{O}_{1}+8i\frac{m_{N}^{2}}{m_{M}^{2}}\mathcal{O}_{5}$
\\
24 & $\bar{\chi}\sigma^{\mu\nu}\chi\bar{N}\left(\gamma_{\mu}\frac{\cancel{q}}{m_{M}}\gamma_{\nu}-\gamma_{\nu}\frac{\cancel{q}}{m_{M}}\gamma_{\mu}\right)N$ & $16i\left(\frac{\vec{\sigma}_{\chi}}{2}\cdot\frac{\vec{q}}{m_{M}}\right)\left(\frac{\vec{\sigma}_{N}}{2}\cdot\vec{v}^{\bot}\right)+\mathrm{\mathit{O}}\left(\frac{1}{m^{3}}\right)$ & $16\frac{m_{N}}{m_{M}}\mathcal{O}_{14}$ 
\\
25 & $\bar{\chi}\left(\frac{q^{\mu}}{m_{M}}\gamma^{\nu}-\frac{q^{\nu}}{m_{M}}\gamma^{\mu}\right)\chi\bar{N}\sigma_{\mu\nu}N$ & $\frac{iq^{2}}{m_{N}m_{M}}1_{\chi}1_{N}+\frac{4}{m_{M}}\frac{\vec{\sigma}_{N}}{2}\cdot\left(\vec{q}\times\vec{v}^{\bot}\right)$ & $i\frac{q^{2}}{m_{N}m_{M}}\mathcal{O}_{1}-4i\frac{m_{N}}{m_{M}}\mathcal{O}_{3}$ 
\\
& & $+\frac{4i}{m_{\chi}m_{M}}\left[q^{2}\left(\frac{\vec{\sigma}_{\chi}}{2}\cdot\frac{\vec{\sigma}_{N}}{2}\right)-\left(\vec{q}\cdot\frac{\vec{\sigma}_{\chi}}{2}\right)\left(\vec{q}\cdot\frac{\vec{\sigma}_{N}}{2}\right)\right]
+\mathit{O}\left(\frac{1}{m^{4}}\right)$ 
& $+4i\frac{q^{2}}{m_{\chi}m_{M}}\mathcal{O}_{4}-4i\frac{m_{N}^{2}}{m_{\chi}m_{M}}\mathcal{O}_{6}$
\\
26 & $\bar{\chi}\left(\frac{q^{\mu}}{m_{M}}\gamma^{\nu}-\frac{q^{\nu}}{m_{M}}\gamma^{\mu}\right)\chi\bar{N}\left(\frac{q_{\mu}}{m_{M}}\gamma_{\nu}-\frac{q_{\nu}}{m_{M}}\gamma_{\mu}\right)N$ 
& $-\frac{iq^{2}}{m_{\chi}m_{M}}1_{\chi}1_{N}
-\frac{4}{m_{M}}\frac{\vec{\sigma}_{\chi}}{2}\cdot\left(\vec{q}\times\vec{v}^{\bot}\right)$
& $-i\frac{q^{2}}{m_{\chi}m_{M}}\mathcal{O}_{1}+4i\frac{m_{N}}{m_{M}}\mathcal{O}_{5}$ 
\\
& & $+\frac{4i}{m_{N}m_{M}}\left[\left(\frac{\vec{\sigma}_{N}}{2}\cdot\vec{q}\right)\left(\frac{\vec{\sigma}_{\chi}}{2}\cdot\vec{q}\right)
-q^{2}\left(\frac{\vec{\sigma}_{\chi}}{2}\cdot\frac{\vec{\sigma}_{N}}{2}\right)\right]
+\mathit{O}\left(\frac{1}{m^{4}}\right)$ 
& $+4i\frac{m_{N}}{m_{M}}\mathcal{O}_{6}-4i\frac{q^{2}}{m_{N}m_{M}}\mathcal{O}_{4}$
\\
27 & $\bar{\chi}\left(\frac{q^{\mu}}{m_{M}}\gamma^{\nu}-\frac{q^{\nu}}{m_{M}}\gamma^{\mu}\right)\chi\bar{N}\left(\frac{q_{\mu}}{m_{M}}\frac{K_{\nu}}{m_{M}}-\frac{q_{\nu}}{m_{M}}\frac{K_{\mu}}{m_{M}}\right)N$ & $-4\frac{m_{N}}{m_{M}}\frac{q^{2}}{m_{M}^{2}}1_{\chi}1_{N}+\mathit{O}\left(\frac{1}{m^{4}}\right)$ & $-4\frac{m_{N}}{m_{M}}\frac{q^{2}}{m_{M}^{2}}\mathcal{O}_{1}$ 
\\
28 & $\bar{\chi}\left(\frac{q^{\mu}}{m_{M}}\frac{P^{\nu}}{m_{M}}-\frac{q^{\nu}}{m_{M}}\frac{P^{\mu}}{m_{M}}\right)\chi\bar{N}\sigma_{\mu\nu}N$ & $2i\frac{m_{\chi}}{m_{N}}\frac{q^{2}}{m_{M}^{2}}1_{\chi}1_{N}+8\frac{m_{\chi}}{m_{M}^{2}}\frac{\vec{\sigma}_{N}}{2}\cdot\left(\vec{q}\times\vec{v}^{\bot}\right)+\mathit{O}\left(\frac{1}{m^{4}}\right)$ & $2i\frac{m_{\chi}}{m_{N}}\frac{q^{2}}{m_{M}^{2}}\mathcal{O}_{1}-8i\frac{m_{\chi}m_{N}}{m_{M}^{2}}\mathcal{O}_{3}$
\\
29 & $\bar{\chi}\left(\frac{q^{\mu}}{m_{M}}\frac{P^{\nu}}{m_{M}}-\frac{q^{\nu}}{m_{M}}\frac{P^{\mu}}{m_{M}}\right)\chi\bar{N}\left(\frac{q_{\mu}}{m_{M}}\gamma_{\nu}-\frac{q_{\nu}}{m_{M}}\gamma_{\mu}\right)N$ & $-4\frac{m_{\chi}}{m_{M}}\frac{q^{2}}{m_{M}^{2}}1_{\chi}1_{N}+\mathit{O}\left(\frac{1}{m^{4}}\right)$ & $-4\frac{m_{\chi}}{m_{M}}\frac{q^{2}}{m_{M}^{2}}\mathcal{O}_{1}$
\\
30 & $\bar{\chi}\left(\frac{q^{\mu}}{m_{M}}\frac{P^{\nu}}{m_{M}}-\frac{q^{\nu}}{m_{M}}\frac{P^{\mu}}{m_{M}}\right)\chi\bar{N}\left(\frac{q_{\mu}}{m_{M}}\frac{K_{\nu}}{m_{M}}-\frac{q_{\nu}}{m_{M}}\frac{K_{\mu}}{m_{M}}\right)N$ & $-8\frac{m_{\chi}m_{N}}{m_{M}^{2}}\frac{q^{2}}{m_{M}^{2}}1_{\chi}1_{N}+\mathit{O}\left(\frac{1}{m^{4}}\right)$ & $-8\frac{m_{\chi}m_{N}}{m_{M}^{2}}\frac{q^{2}}{m_{M}^{2}}\mathcal{O}_{1}$ 
\\
31 & $\bar{\chi}\left(\gamma^{\mu}\frac{\cancel{q}}{m_{M}}\gamma^{\nu}-\gamma^{\nu}\frac{\cancel{q}}{m_{M}}\gamma^{\mu}\right)\chi\bar{N}\sigma_{\mu\nu}N$ & $-16i\left(\frac{\vec{\sigma}_{N}}{2}\cdot\frac{\vec{q}}{m_{M}}\right)\left(\frac{\vec{\sigma}_{\chi}}{2}\cdot\vec{v}^{\bot}\right)+\mathrm{\mathit{O}}\left(\frac{1}{m^{4}}\right)$ & $-16\frac{m_{N}}{m_{M}}\mathcal{O}_{13}$ 
\\
32 & $\bar{\chi}\left(\gamma^{\mu}\frac{\cancel{q}}{m_{M}}\gamma^{\nu}-\gamma^{\nu}\frac{\cancel{q}}{m_{M}}\gamma^{\mu}\right)\chi\bar{N}\left(\gamma_{\mu}\frac{\cancel{q}}{m_{M}}\gamma_{\nu}-\gamma_{\nu}\frac{\cancel{q}}{m_{M}}\gamma_{\mu}\right)N$ & $\frac{32}{m_{M}^{2}}\left[q^{2}\left(\frac{\vec{\sigma}_{\chi}}{2}\cdot\frac{\vec{\sigma}_{N}}{2}\right)-\left(\frac{\vec{\sigma}_{N}}{2}\cdot\vec{q}\right)\left(\frac{\vec{\sigma}_{\chi}}{2}\cdot\vec{q}\right)\right]+\mathrm{\mathit{O}}\left(\frac{1}{m^{4}}\right)$ & $32\frac{q^{2}}{m_{M}^{2}}\mathcal{O}_{4}-32\frac{m_{N}^{2}}{m_{M}^{2}}\mathcal{O}_{6}$
\\
33 & $\bar{\chi}\sigma^{\mu\nu}\chi\bar{N}\sigma_{\mu\nu}\gamma_{5}N$ & $-2\left(\frac{\vec{\sigma}_{N}}{2}\cdot\frac{\vec{q}}{m_{\chi}}\right)+2\left(\frac{\vec{\sigma}_{\chi}}{2}\cdot\frac{\vec{q}}{m_{N}}\right)$ & $2i\frac{m_{N}}{m_{\chi}}\mathcal{O}_{10}-2i\mathcal{O}_{11}$
\\
 &  & $+8i\frac{\vec{\sigma}_{\chi}}{2}\cdot\left(\frac{\vec{\sigma}_{N}}{2}\times\vec{v}^{\bot}\right)+\mathrm{\mathit{O}}\left(\frac{1}{m^{3}}\right)$ & $+8i\mathcal{O}_{12}$
 \\
34 & $\bar{\chi}\sigma^{\mu\nu}\chi\bar{N}\left(\frac{q_{\mu}}{m_{M}}\gamma_{\nu}-\frac{q_{\nu}}{m_{M}}\gamma_{\mu}\right)\gamma_{5}N$ & $-8\frac{\vec{\sigma}_{\chi}}{2}\cdot\left(\frac{\vec{\sigma}_{N}}{2}\times\frac{\vec{q}}{m_{M}}\right)+\mathit{O}\left(\frac{1}{m^{3}}\right)$ & $8i\frac{m_{N}}{m_{M}}\mathcal{O}_{9}$
\\
35 & $\bar{\chi}\sigma^{\mu\nu}\chi\bar{N}\left(\frac{q_{\mu}}{m_{M}}\frac{K_{\nu}}{m_{M}}-\frac{q_{\nu}}{m_{M}}\frac{K_{\mu}}{m_{M}}\right)\gamma_{5}N$ & $-2i\frac{q^{2}}{m_{M}^{2}}\left(\frac{\vec{q}}{m_{\chi}}\cdot\frac{\vec{\sigma}_{N}}{2}\right)+8\left(\frac{\vec{\sigma}_{\chi}}{2}\times\vec{v}^{\bot}\right)\cdot\frac{\vec{q}}{m_{M}}\left(\frac{\vec{\sigma}_{N}}{2}\cdot\frac{\vec{q}}{m_{M}}\right)+\mathit{O}\left(\frac{1}{m^{5}}\right)$ & $-2\frac{m_{N}}{m_{\chi}}\frac{q^{2}}{m_{M}^{2}}\mathcal{O}_{10}-8\frac{m_{N}^{2}}{m_{M}^{2}}\mathcal{O}_{16}$
\\
36 & $\bar{\chi}\sigma^{\mu\nu}\chi\bar{N}\left(\gamma_{\mu}\frac{\cancel{q}}{m_{M}}\gamma_{\nu}-\gamma_{\nu}\frac{\cancel{q}}{m_{M}}\gamma_{\mu}\right)\gamma_{5}N$ & $-8\frac{\vec{\sigma}_{\chi}}{2}\cdot\frac{\vec{q}}{m_{M}}+\mathrm{\mathit{O}}\left(\frac{1}{m^{3}}\right)$ & $8i\frac{m_{N}}{m_{M}}\mathcal{O}_{11}$
\\
37 & $\bar{\chi}\left(\frac{q^{\mu}}{m_{M}}\gamma^{\nu}-\frac{q^{\nu}}{m_{M}}\gamma^{\mu}\right)\chi\bar{N}\sigma_{\mu\nu}\gamma_{5}N$ & $4i\frac{\vec{q}}{m_{M}}\cdot\frac{\vec{\sigma}_{N}}{2}+\mathrm{\mathit{O}}\left(\frac{1}{m^{3}}\right)$ & $4\frac{m_{N}}{m_{M}}\mathcal{O}_{10}$
\\
38 & $\bar{\chi}\left(\frac{q^{\mu}}{m_{M}}\gamma^{\nu}-\frac{q^{\nu}}{m_{M}}\gamma^{\mu}\right)\chi\bar{N}\left(\frac{q_{\mu}}{m_{M}}\gamma_{\nu}-\frac{q_{\nu}}{m_{M}}\gamma_{\mu}\right)\gamma_{5}N$ & $4\frac{q^{2}}{m_{M}^{2}}\left[\left(\vec{v}^{\bot}\cdot\frac{\vec{\sigma}_{N}}{2}\right)-i\frac{\vec{\sigma}_{\chi}}{2}\cdot\left(\frac{\vec{\sigma}_{N}}{2}\times\frac{\vec{q}}{m_{\chi}}\right)\right]+\mathit{O}\left(\frac{1}{m^{5}}\right)$ & $4\frac{q^{2}}{m_{M}^{2}}\left(\mathcal{O}_{7}-\frac{m_{N}}{m_{\chi}}\mathcal{O}_{9}\right)$
\\
39 & $\bar{\chi}\left(\frac{q^{\mu}}{m_{M}}\gamma^{\nu}-\frac{q^{\nu}}{m_{M}}\gamma^{\mu}\right)\chi\bar{N}\left(\frac{q_{\mu}}{m_{M}}\frac{K_{\nu}}{m_{M}}-\frac{q_{\nu}}{m_{M}}\frac{K_{\mu}}{m_{M}}\right)\gamma_{5}N$ & $-2\frac{q^{2}}{m_{M}^{2}}\left(\frac{\vec{\sigma}_{N}}{2}\cdot\frac{\vec{q}}{m_{M}}\right)+\mathit{O}\left(\frac{1}{m^{5}}\right)$ & $2i\frac{m_{N}}{m_{M}}\frac{q^{2}}{m_{M}^{2}}\mathcal{O}_{10}$
\\
40 & $\bar{\chi}\left(\frac{q^{\mu}}{m_{M}}\frac{P^{\nu}}{m_{M}}-\frac{q^{\nu}}{m_{M}}\frac{P^{\mu}}{m_{M}}\right)\chi\bar{N}\sigma_{\mu\nu}\gamma_{5}N$ & $8i\frac{m_{\chi}}{m_{M}}\left(\frac{\vec{q}}{m_{M}}\cdot\frac{\vec{\sigma}_{N}}{2}\right)+\mathit{O}\left(\frac{1}{m^{3}}\right)$ & $8\frac{m_{\chi}m_{N}}{m_{M}^{2}}\mathcal{O}_{10}$
\\
41 & $\bar{\chi}\left(\frac{q^{\mu}}{m_{M}}\frac{P^{\nu}}{m_{M}}-\frac{q^{\nu}}{m_{M}}\frac{P^{\mu}}{m_{M}}\right)\chi\bar{N}\left(\frac{q_{\mu}}{m_{M}}\gamma_{\nu}-\frac{q_{\nu}}{m_{M}}\gamma_{\mu}\right)\gamma_{5}N$ & $8\frac{m_{\chi}}{m_{M}}\frac{q^{2}}{m_{M}^{2}}\left(\vec{v}^{\bot}\cdot\frac{\vec{\sigma}_{N}}{2}\right)+\mathit{O}\left(\frac{1}{m^{5}}\right)$ & $8\frac{m_{\chi}}{m_{M}}\frac{q^{2}}{m_{M}^{2}}\mathcal{O}_{7}$
\\
42 & $\bar{\chi}\left(\frac{q^{\mu}}{m_{M}}\frac{P^{\nu}}{m_{M}}-\frac{q^{\nu}}{m_{M}}\frac{P^{\mu}}{m_{M}}\right)\chi\bar{N}\left(\frac{q_{\mu}}{m_{M}}\frac{K_{\nu}}{m_{M}}-\frac{q_{\nu}}{m_{M}}\frac{K_{\mu}}{m_{M}}\right)\gamma_{5}N$ & $-8\frac{m_{\chi}}{m_{M}}\frac{q^{2}}{m_{M}^{2}}\left(\frac{\vec{q}}{m_{M}}\cdot\frac{\vec{\sigma}_{N}}{2}\right)+\mathit{O}\left(\frac{1}{m^{5}}\right)$ & $8i\frac{m_{\chi}m_{N}}{m_{M}^{2}}\frac{q^{2}}{m_{M}^{2}}\mathcal{O}_{10}$
\\
43 & $\bar{\chi}\left(\gamma^{\mu}\frac{\cancel{q}}{m_{M}}\gamma^{\nu}-\gamma^{\nu}\frac{\cancel{q}}{m_{M}}\gamma^{\mu}\right)\chi\bar{N}\sigma_{\mu\nu}\gamma_{5}N$ & $-8\frac{\vec{\sigma}_{\chi}}{2}\cdot\left(\frac{\vec{\sigma}_{N}}{2}\times\frac{\vec{q}}{m_{M}}\right)+\mathrm{\mathit{O}}\left(\frac{1}{m^{3}}\right)$ & $8i\frac{m_{N}}{m_{M}}\mathcal{O}_{9}$
\\
44 & $\bar{\chi}\left(\gamma^{\mu}\frac{\cancel{q}}{m_{M}}\gamma^{\nu}-\gamma^{\nu}\frac{\cancel{q}}{m_{M}}\gamma^{\mu}\right)\chi\bar{N}\left(\gamma_{\mu}\frac{\cancel{q}}{m_{M}}\gamma_{\nu}-\gamma_{\nu}\frac{\cancel{q}}{m_{M}}\gamma_{\mu}\right)\gamma_{5}N$ & $-16\frac{q^{2}}{m_{M}^{2}}\left[\left(\vec{v}^{\bot}\cdot\frac{\vec{\sigma}_{\chi}}{2}\right)-i\frac{\vec{\sigma}_{\chi}}{2}\cdot\left(\frac{\vec{\sigma}_{N}}{2}\times\frac{\vec{q}}{m_{N}}\right)\right]+\mathrm{\mathit{O}}\left(\frac{1}{m^{5}}\right)$ & $-16\frac{q^{2}}{m_{M}^{2}}\left(\mathcal{O}_{8}-\mathcal{O}_{9}\right)$
\end{tabular}
\end{ruledtabular}
  \end{minipage}
  }
  \end{table*}

To incorporate the missing tensor symmetry into the framework, we write the interaction of the DM and the nucleons using all the possible Lorentz-invariant tensor terms, similar to how non-tensor terms were treated~\cite{fitzpatrick2013effective, anand2014weakly}.
We conclude them based on the general form of the single-nucleon matrix element between nuclear states of the tensor part of the nuclear current~\cite{Winberg1958charge, Cirigliano2013beta}:
\begin{multline}
\left\langle k_f\left|
\bar{q}
\sigma_{\mu \nu}\tau^a q\right|k_i\right\rangle\\
= \bar{u}\left(k_f\right)\left[g_{T}\left(q^{2}\right)\sigma_{\mu\nu}
 +\tilde{g}_{T}^{\left(1\right)}\left(q^{2}\right)\left(\frac{q_{\mu}}{m_{M}}\gamma_{\nu}-\frac{q_{\nu}}{m_{M}}\gamma_{\mu}\right)\right.\\
+\tilde{g}_{T}^{\left(2\right)}\left(q^{2}\right)\left(\frac{q_{\mu}}{m_{M}}\frac{K_{\nu}}{m_{M}}-\frac{q_{\nu}}{m_{M}}\frac{K_{\mu}}{m_{M}}\right)\\
 \left.+\tilde{g}_{T}^{\left(3\right)}\left(q^{2}\right)\left(\gamma_{\mu}\frac{\cancel{q}}{m_{M}}\gamma_{\nu}-\gamma_{\nu}\frac{\cancel{q}}{m_{M}}\gamma_{\mu}\right)\right]
 \tau^a u\left(k_i\right)\mbox{,}
  \label{eq: tensor_current}
\end{multline}
which consists of all the possible Lorentz-invariant tensor combinations~\footnote{The last term is absent in some of the literature (see, e.g.,~\cite{adler1975renormalization,Hoferichter2019Nucleon}) due to its classification as a second-class current which vanishes in the isospin limit~\cite{adler1975renormalization,Cirigliano2013beta}.
However, due to the lack of specific knowledge regarding the interaction, this analysis takes a comprehensive approach to ensure we do not overlook any possible contributions.}.
Here we use the nowadays convention for the gamma matrices $\gamma_{\mu}$ and their commutator $\sigma_{\mu\nu}$ (see, e.g.,~\cite{itzykson1980quantum}).
The form factors $\tilde{g}_T^{\left(i\right)}$ ($i\in\left\{1, 2, 3 \right\}$) are defined as the dimensionless version of the standard tensor form factors $g_T^{\left(i\right)}$ given in~\cite{Cirigliano2013beta}. 
\footnote{We introduce the redefined form factors as follows: 
${g}_T^{\left(1\right)}=\frac{\Tilde{g}_T^{\left(1\right)}}{m_M}$, ${g}_T^{\left(2\right)}=\frac{\Tilde{g}_T^{\left(2\right)}}{m^2_M}$, and ${g}_T^{\left(3\right)}=\frac{\Tilde{g}_T^{\left(3\right)}}{m_M}$, where $m_M$ is the relevant theory-dependent mass scale (e.g., the nucleon mass) given a model context.
This adjustment was made to align this work with~\cite{anand2014weakly}, where similar theory-dependent masses were utilized.}
We assume a similar matrix element for the WIMP tensor current between the initial and final WIMP states, differing only in the form factors, and exclude the isospin operator $\tau ^a$ ($a\in\left\{1, 2, 3 \right\}$).

Given the presence of four Lorentz-invariant tensor terms in the current, there exist 16 potential combinations for the tensor coupling between the nuclear and WIMPs currents, but 4 of them will vanish, leaving us with only 12 combinations.
As done by Lee and Yang for the weak interaction~\cite{lee1956question}, to these 12 basic combinations we add their $\gamma_5$ variations, resulting in 24 new combinations.
Utilizing the antisymmetric tensor-current decomposition~\cite{glick2023multipole}:
\begin{equation}
\begin{split}
\vec{\mathcal{J}}^{T}_{i} & \equiv -\frac{i}{\sqrt{2}}\epsilon_{ijk}\mathcal{J}_{jk},\label{eq: tensor current defenition T} \\
\vec{\mathcal{J}}^{T'}_{i} & \equiv \sqrt{2}\mathcal{J}_{0i}\text{,}
\end{split}
\end{equation}
and applying the following identity for each combination of two tensor-terms coupling:
\begin{eqnarray}
\label{eq:vectors products}
j_{\mu\nu}J^{\mu\nu} & = & -\left[\vec{j}^{T}\cdot\vec{J}^{T}+\vec{j}^{T'}\cdot\vec{J}^{T'}\right]\mbox{,}
\end{eqnarray}
%
we achieve a nonrelativistic reduction in leading orders in $1/\text{mass}$ for each combination. These reductions, listed in table~\ref{tab: WIMP Nonrelativistic Reduction}, give rise to all the nonrelativistic operators already existing in the scalar and vector cases~\cite{anand2014weakly}:
\begin{equation}
\begin{split}
\mathcal{O}_{1} & \equiv 1_{\chi}1_{N} \text{,}\\
\mathcal{O}_{3} & \equiv i\vec{S}_{N}\cdot\left(\frac{\vec{q}}{m_{N}}\times\vec{v}^{\bot}\right) \text{,}\\
\mathcal{O}_{4} & \equiv \vec{S}_{\chi}\cdot\vec{S}_{N} \text{,}\\
\mathcal{O}_{5} & \equiv i\vec{S}_{\chi}\cdot\left(\frac{\vec{q}}{m_{N}}\times\vec{v}^{\bot}\right) \text{,}\\
\mathcal{O}_{6} & \equiv \left(\vec{S}_{\chi}\cdot\frac{\vec{q}}{m_{N}}\right)\left(\vec{S}_{N}\cdot\frac{\vec{q}}{m_{N}}\right) \text{,}\\
\mathcal{O}_{7} & \equiv \vec{S}_{N}\cdot \vec{v}^{\bot} \text{,}\\
\mathcal{O}_{8} & \equiv \vec{S}_{\chi}\cdot \vec{v}^{\bot} \text{,}\\
\mathcal{O}_{9} & \equiv i\vec{S}_{\chi}\cdot\left(\vec{S}_{N}\times\frac{\vec{q}}{m_{N}}\right) \text{,}\\
\mathcal{O}_{10} & \equiv i\vec{S}_{N}\cdot\frac{\vec{q}}{m_{N}} \text{,}\\
\mathcal{O}_{11} & \equiv i\vec{S}_{\chi}\cdot\frac{\vec{q}}{m_{N}} \text{,}\\
\mathcal{O}_{12} & \equiv \vec{S}_{\chi}\cdot\left(\vec{S}_{N}\times\vec{v}^{\bot}\right) \text{,}\\
\mathcal{O}_{13} & \equiv i\left(\vec{S}_{\chi}\cdot\vec{v}^{\bot}\right)\left(\vec{S}_{N}\cdot\frac{\vec{q}}{m_{N}}\right) \text{,}\\
\mathcal{O}_{14} & \equiv i\left(\vec{S}_{\chi}\cdot\frac{\vec{q}}{m_{N}}\right)\left(\vec{S}_{N}\cdot\vec{v}^{\bot}\right) \text{,}\\
\mathcal{O}_{15} & \equiv -\left(\vec{S}_{\chi}\cdot\frac{\vec{q}}{m_{N}}\right)\left[\left(\vec{S}_{N}\times\vec{v}^{\bot}\right) \cdot \frac{\vec{q}}{m_{N}} \right] \text{,}\\
\mathcal{O}_{16} & \equiv -\left[\left(\vec{S}_{\chi}\times\vec{v}^{\bot}\right) \cdot \frac{\vec{q}}{m_{N}} \right]\left(\vec{S}_{N}\cdot\frac{\vec{q}}{m_{N}}\right) 
\end{split}
\label{eq: NRET_operators_WIMP}
\end{equation}
(note the dependence $\mathcal{O}_{16} = \mathcal{O}_{15}+\frac{q^2}{m_N^2}\mathcal{O}_{12}$).

Upon detecting DM scattering off nuclei, the subsequent step involves scrutinizing the specific nature of the measured interaction between DM and nucleons.
It is essential to note that each set of terms sharing the same WIMP current part and $\gamma_5$ coupling (21-24, 25-27, 28-30, 31-32, 33-36, 37-39, 40-42, and 43-44 in Table~\ref{tab: WIMP Nonrelativistic Reduction}) should possess the same new-physics form factor, multiplied by the corresponding nucleon tensor form-factor $g_T$ or $\tilde{g}_T^{\left(i\right)}$, contingent on the nuclear current part as outlined in Eq.~\eqref{eq: tensor_current}.
Calculations for nucleon tensor form-factors are available, see~\cite{Gonzalez2019progress} and references within for $g_T$, and, e.g.,~\cite{Hoferichter2019Nucleon}, for the other form factors. 
So despite the seemingly numerous terms, the tensor WIMP-nucleon interaction introduces only four plus four (the regular tensor couplings plus the $\gamma_5$ tensor couplings) new-physics coefficients for the measurements to constrain.

Identifying the operators associated with each symmetry case is pivotal for discerning the nature of the WIMP-nucleon interaction and determining its symmetries. Once an operator aligns with measured data, knowing all the symmetries it is involved in becomes imperative. Therefore, although the tensor terms do not introduce new operators, their operator reduction is necessary to establish whether the WIMP-nucleon interaction consists of tensor symmetry.
\paragraph*{$\mu \to e$ conversion.\label{sec: mu-e}}
Unlike the interaction of DM with nuclei, elastic $\mu \to e$ conversion involves certain tensor terms that encompass operators absent from non-tensor cases, as emphasized in~\cite{Rule2023nuclear, haxton2023nuclear}. Following the NRET framework for $\mu \to e$ conversion as constructed in the references above, we employ the lepton (nucleon) identity operators $1_L$ ($1_N$), along with five dimensionless Hermitian three-vectors: $i\hat{q}$, where $\hat{q}$ is the unit vector along the three-momentum transfer to the leptons; $\vec{v}_N \equiv \frac{\vec{p}_i + \vec{p}_f}{2m_N}$, representing the nuclear velocity and symmetrically combining the initial and final nucleon velocities $\frac{\vec{p}_i}{m_N}$ and $\frac{\vec{p}_f}{m_N}$; $\vec{v}_{\mu}$, denoting the muon velocity relative to the center-of-mass of the nucleons; and $\vec{\sigma}_L$ and $\vec{\sigma}_N$, representing the spins of the leptons and the nucleons, respectively.

Following the NRET that has been constructed in~\cite{Rule2023nuclear, haxton2023nuclear} for $\mu \to e$ conversion,
we employ lepton (nucleon) identity operators $1_L$ ($1_N$), along with five dimensionless Hermitian three-vectors: $i\hat{q}$, where $\hat{q}$ represents the unit vector along the three-momentum transfer to the leptons, $\vec{v}_N\equiv\frac{\vec{p}_i+\vec{p}_f}{2m_N}$ denotes the nuclear velocity, symmetrically combining the initial and final nucleon velocities $\frac{\vec{p}_i}{m_N}$ and $\frac{\vec{p}_f}{m_N}$, $\vec{v}_{\mu}$ stands for the muon velocity relative to the center-of-mass of the nucleons, and $\vec{\sigma}_L$ and $\vec{\sigma}_N$ represent the spins of the leptons and the nucleons, respectively.

Conducting a parallel reduction to that performed for WIMPs scattering, we commence from the same four possible tensor Lorentz-invariant terms presented in Eq.~\eqref{eq: tensor_current}.
These are all the tensor possible contact terms (refer to~\cite{dekens2019non} for further discussion on non-contact terms). In the Standard Model effective field theory (SMEFT) Lagrangian, a singular tensor term exists~\cite{buchmuller1986effective, grzadkowski2010dimension}:
\begin{align}
\mathcal{\hat{L}}_{\text{SMEFT}} & \supset
\frac{c^{\alpha \beta m n}}{\Lambda_{\text{BSM}}^2}\bar{l}^{J\alpha}_L\sigma_{\mu \nu} e_R^{\beta}\epsilon_{JK}\bar{q}^{K m}_L \sigma^{\mu \nu} u
_R^n.
\end{align}
Here, $\Lambda_{\text{BSM}}$ denotes the new-physics scale, $c$ represents the coefficient of this tensor term, and the indexes $\alpha, \beta, m, n \in \left\{1, 2, 3\right\}$ represent the generations of the particles (further details and conventions can be found in~\cite{Cirigliano2013beta}). Applying the left lepton doublet
$l_L^{\alpha}=\begin{pmatrix}
\nu_L^{\alpha}\\
e_L^{\alpha}
\end{pmatrix}$,
the left quark doublet
$q_L^m=\begin{pmatrix}
u_L^m\\
d_L^m
\end{pmatrix}$,
and the antisymmetric two-dimensional matrix
$\epsilon=\begin{pmatrix}
0 & 1\\
-1 & 0
\end{pmatrix}$
yields two terms:
\begin{align}
\bar{\nu}_L^{\alpha}\sigma_{\mu \nu} e_R^{\beta}\bar{d}^m_L \sigma^{\mu \nu} u_R^n
- \bar{e}_L^{\alpha}\sigma_{\mu \nu} e_R^{\beta} \bar{u}^m_L \sigma^{\mu \nu} u_R^n.
\end{align}
The first illustrates quark isospin exchange processes (e.g., $\beta$-decays),
while the second preserves quark isospin and can lead to $\mu \to e$ conversion when taking
$\alpha=1$ and $\beta=2$,
i.e.,
\begin{align}
- \bar{e}_L \sigma_{\mu \nu} \mu_R \bar{u}^m_L \sigma^{\mu \nu} u_R^n.
\end{align}
Notably, this implies that $\mu \to e$ conversion can only occur with up, charm, or top quarks (for $m=n=1$, $2$, or $3$, respectively).
\begin{table*}
\centering
\rotatebox{90}{
\begin{minipage}{\textheight}
  \caption{\label{tab: mu-to-e Nonrelativistic Reduction}The second column showcases the tensor Lagrangian densities $\mathcal{L}_{\text{int}}^{j}$ corresponding to elastic $\mu \to e$ conversion, with the index $j$ enumerated in the first column. The terms encompass scalar and vector contributions outlined in \cite{haxton2023nuclear} ($j \in \left\{1, 2, ..., 20\right\}$), augmented here by the inclusion of previously unaccounted tensor terms. In the third column, the operators resulting from a nonrelativistic reduction between Pauli spinors are presented, while the fourth column illustrates the associated effective interactions expressed in terms of the NRET operators defined in Eqs. \eqref{eq: NRET_operators_mu} and \eqref{eq: NRET_operators_mu_f}. Further details can be found in \cite{haxton2023nuclear}.}
\begin{ruledtabular}
\begin{tabular}{cccc}
$j$ & $\mathcal{L}_{\text{int}}^{j}$ & Pauli operator reduction & $\Sigma_{i}c_{i}\mathcal{O}_{i}$
\\
\hline 
21 & $\bar{\chi}_{e}\sigma^{\mu\nu}\chi_{\mu}\bar{N}\sigma_{\mu\nu}N$ & $-\frac{q}{m_{N}}1_{L}1_{N}-2i1_{L}\hat{q}\cdot\left(\vec{v}_{N}\times\vec{\sigma}_{N}\right)+2\vec{\sigma}_{L}\cdot\vec{\sigma}_{N}+2\vec{\sigma}_{L}\cdot\left[\hat{q}\times\left(\vec{v}_{N}\times\vec{\sigma}_{N}\right)\right]$ & $-\frac{q}{m_{N}}\mathcal{O}_{1}-2\mathcal{O}_{3}+2\mathcal{O}_{4}-2i\mathcal{O}_{13}^{'}$
\\
 &  & $+i\left(\hat{q}\times\vec{v}_{\mu}\right)\cdot\vec{\sigma}_{N}-\left(\vec{v}_{\mu}\cdot\vec{\sigma}_{L}\right)\left(\hat{q}\cdot\vec{\sigma}_{N}\right)-\left[\hat{q}\times\left(\vec{v}_{\mu}\times\vec{\sigma}_{L}\right)\right]\cdot\vec{\sigma}_{N}+\mathit{O}\left(\frac{q^{2}}{m_{N}^{2}}\right)$ & $+2\mathcal{O}_{5}^{f}+2i\mathcal{O}_{14}^{f}+2i\mathcal{O}_{13}^{f'}$ 
 \\
22 & $\bar{\chi}_{e}\sigma^{\mu\nu}\chi_{\mu}\bar{N}\left(\frac{q_{\mu}}{m_{N}}\gamma_{\nu}-\frac{q_{\nu}}{m_{N}}\gamma_{\mu}\right)N$ & $\frac{q}{m_{N}}\left[-2i1_{L}1_{N}+i\frac{q}{m_{N}}\left(\vec{\sigma}_{L}\cdot\vec{\sigma}_{N}\right)+i\frac{q}{m_{N}}\left(\hat{q}\cdot\vec{\sigma}_{N}\right)\left(\vec{\sigma}_{L}\cdot\hat{q}\right)\right.$ & $i\frac{q}{m_{N}}\left(-2\mathcal{O}_{1}+\frac{q}{m_{N}}\mathcal{O}_{4}-\frac{q}{m_{N}}\mathcal{O}_{6}\right.$   \\
 &  & $\left.-2\vec{\sigma}_{L}\cdot\left(\hat{q}\times\vec{v}_{N}\right)+i\left(\vec{v}_{\mu}\cdot\hat{q}\right)1_{N}-\hat{q}\cdot\left(\vec{v}_{\mu}\times\vec{\sigma}_{L}\right)1_{N}\right]+\mathit{O}\left(\frac{q^{3}}{m_{N}^{3}}\right)$ & $\left.+2\mathcal{O}_{5}-2i\mathcal{O}_{2}^{f'}+2\mathcal{O}_{3}^{f}\right)$   \\
23 & $\bar{\chi}_{e}\sigma^{\mu\nu}\chi_{\mu}\bar{N}\left(\frac{q_{\mu}}{m_{N}}v_{N\nu}-\frac{q_{\nu}}{m_{N}}v_{N\mu}\right)N$ & $\frac{q}{m_{N}}\left[-2i1_{L}\cdot1_{N}+2\vec{\sigma}_{L}\cdot\left(\hat{q}\times\vec{v}_{N}\right)+i\left(\hat{q}\cdot\vec{v}_{\mu}\right)1_{N}-\hat{q}\cdot\left(\vec{v}_{\mu}\times\vec{\sigma}_{L}\right)1_{N}\right]+\mathit{O}\left(\frac{q^{3}}{m_{N}^{3}}\right)$ & $2\frac{q}{m_{N}}\left(-i\mathcal{O}_{1}-i\mathcal{O}_{5}+\mathcal{O}_{2}^{f'}+i\mathcal{O}_{3}^{f}\right)$   \\
24 & $\bar{\chi}_{e}\sigma^{\mu\nu}\chi_{\mu}\bar{N}\left(\gamma_{\mu}\frac{\cancel{q}}{m_{N}}\gamma_{\nu}-\gamma_{\nu}\frac{\cancel{q}}{m_{N}}\gamma_{\mu}\right)N$ & $-4i\frac{q}{m_{N}}\left\{ \left(\vec{\sigma}_{L}\cdot\vec{\sigma}_{N}\right)-\left(\vec{\sigma}_{L}\cdot\hat{q}\right)\left(\vec{\sigma}_{N}\cdot\hat{q}\right)+\left(\hat{q}\cdot\vec{\sigma}_{L}\right)\left(\vec{v}_{N}\cdot\vec{\sigma}_{N}\right)\right.$ & $-4i\frac{q}{m_{N}}\left(\mathcal{O}_{4}+\mathcal{O}_{6}-i\mathcal{O}_{14}\right.$   \\
 &  & $\left.+i\left(\hat{q}\times\frac{\vec{v}_{\mu}}{2}\right)\cdot\vec{\sigma}_{N}-\left[\hat{q}\times\left(\frac{\vec{v}_{\mu}}{2}\times\vec{\sigma}_{L}\right)\right]\cdot\vec{\sigma}_{N}\right\} +\mathit{O}\left(\frac{q^{3}}{m_{N}^{3}}\right)$ & $\left.+\mathcal{O}_{5}^{f}+i\mathcal{O}_{13}^{f'}\right)$   \\
25 & $\bar{\chi}_{e}\left(\hat{q}^{\mu}\gamma^{\nu}-\hat{q}^{\nu}\gamma^{\mu}\right)\chi_{\mu}\bar{N}\sigma_{\mu\nu}N$ & $2i\left(\vec{\sigma}_{L}\cdot\vec{\sigma}_{N}\right)-2i\left(\hat{q}\cdot\vec{\sigma}_{L}\right)\left(\hat{q}\cdot\vec{\sigma}_{N}\right)+\frac{i}{2}\frac{q}{m_{N}}1_{L}1_{N}-1_{L}\hat{q}\cdot\left(\vec{v}_{N}\times\vec{\sigma}_{N}\right)$ & $i\left(2\mathcal{O}_{4}+2\mathcal{O}_{6}+\frac{1}{2}\frac{q}{m_{N}}\mathcal{O}_{1}+\mathcal{O}_{3}\right.$   \\
 &  & $+\left(\hat{q}\times\vec{v}_{\mu}\right)\cdot\vec{\sigma}_{N}+i\left[\hat{q}\times\left(\vec{v}_{\mu}\times\vec{\sigma}_{L}\right)\right]\cdot\vec{\sigma}_{N}+\mathit{O}\left(\frac{q^{2}}{m_{N}^{2}}\right)$ & $\left.-2\mathcal{O}_{5}^{f}-2i\mathcal{O}_{13}^{f'}\right)$   \\
26 & $\bar{\chi}_{e}\left(\hat{q}^{\mu}\gamma^{\nu}-\hat{q}^{\nu}\gamma^{\mu}\right)\chi_{\mu}\bar{N}\left(\frac{q_{\mu}}{m_{N}}\gamma_{\nu}-\frac{q_{\nu}}{m_{N}}\gamma_{\mu}\right)N$ & $\frac{q}{m_{N}}\left[-1_{L}1_{N}+\frac{q}{m_{N}}\left(\vec{\sigma}_{L}\cdot\vec{\sigma}_{N}\right)-\frac{q}{m_{N}}\left(\hat{q}\cdot\vec{\sigma}_{L}\right)\left(\hat{q}\cdot\vec{\sigma}_{N}\right)\right.$ & $\frac{q}{m_{N}}\left(-\mathcal{O}_{1}+\frac{q}{m_{N}}\mathcal{O}_{4}+\frac{q}{m_{N}}\mathcal{O}_{6}\right.$   \\
 &  & $\left.+2i\vec{\sigma}_{L}\cdot\left(\hat{q}\times\vec{v}_{N}\right)-\left(\hat{q}\cdot\vec{v}_{\mu}\right)1_{N}-i\hat{q}\cdot\left(\vec{v}_{\mu}\times\vec{\sigma}_{L}\right)1_{N}\right]+\mathit{O}\left(\frac{q^{3}}{m_{N}^{3}}\right)$ & $\left.+2\mathcal{O}_{5}+2i\mathcal{O}_{2}^{f'}-2\mathcal{O}_{3}^{f}\right)$   \\
27 & $\bar{\chi}_{e}\left(\hat{q}^{\mu}\gamma^{\nu}-\hat{q}^{\nu}\gamma^{\mu}\right)\chi_{\mu}\bar{N}\left(\frac{q_{\mu}}{m_{N}}v_{N\nu}-\frac{q_{\nu}}{m_{N}}v_{N\mu}\right)N$ & $\frac{q}{m_{N}}\left[-1_{L}1_{N}+2i\vec{\sigma}_{L}\cdot\left(\hat{q}\times\vec{v}_{N}\right)\right.$ & $\frac{q}{m_{N}}\left(-\mathcal{O}_{1}+2\mathcal{O}_{5}\right.$   \\
 &  & $\left.-\left(\hat{q}\cdot\vec{v}_{\mu}\right)1_{N}-i\hat{q}\cdot\left(\vec{v}_{\mu}\times\vec{\sigma}_{L}\right)1_{N}\right]+\mathit{O}\left(\frac{q^{3}}{m_{N}^{3}}\right)$ & $\left.+2i\mathcal{O}_{2}^{f'}-2\mathcal{O}_{3}^{f}\right)$   \\
28 & $\bar{\chi}_{e}\left(\hat{q}^{\alpha}v_{\mu}^{\nu}-\hat{q}^{\nu}v_{\mu}^{\alpha}\right)\chi_{\mu}\bar{N}\sigma_{\alpha\nu}N$ & $2\left[\frac{i}{2}\frac{q}{m_{N}}1_{L}1_{N}-1_{L}\hat{q}\cdot\left(\vec{v}_{N}\times\vec{\sigma}_{N}\right)+\left(\hat{q}\times\vec{v}_{\mu}\right)\cdot\vec{\sigma}_{N}\right]+\mathrm{\mathit{O}}\left(\frac{q^{2}}{m_{N}^{2}}\right)$ & $i\left(\frac{q}{m_{N}}\mathcal{O}_{1}+2\mathcal{O}_{3}-4\mathcal{O}_{5}^{f}\right)$   \\
29 & $\bar{\chi}_{e}\left(\hat{q}^{\alpha}v_{\mu}^{\nu}-\hat{q}^{\nu}v_{\mu}^{\alpha}\right)\chi_{\mu}\bar{N}\left(\frac{q_{\alpha}}{m_{N}}\gamma_{\nu}-\frac{q_{\nu}}{m_{N}}\gamma_{\alpha}\right)N$ & $2\frac{q}{m_{N}}\left[-1_{L}1_{N}+\left(\hat{q}\cdot\frac{\vec{v}_{\mu}}{2}\right)1_{N}+i\hat{q}\cdot\left(\frac{\vec{v}_{\mu}}{2}\times\vec{\sigma}_{L}\right)1_{N}\right]+\mathit{O}\left(\frac{q^{3}}{m_{N}^{3}}\right)$ & $2\frac{q}{m_{N}}\left(-\mathcal{O}_{1}-i\mathcal{O}_{2}^{f'}+\mathcal{O}_{3}^{f}\right)$   \\
30 & $\bar{\chi}_{e}\left(\hat{q}^{\alpha}v_{\mu}^{\nu}-\hat{q}^{\nu}v_{\mu}^{\alpha}\right)\chi_{\mu}\bar{N}\left(\frac{q_{\alpha}}{m_{N}}v_{N\nu}-\frac{q_{\nu}}{m_{N}}v_{N\alpha}\right)N$ & $2\frac{q}{m_{N}}\left[-1_{L}1_{N}+\left(\hat{q}\cdot\frac{\vec{v}_{\mu}}{2}\right)1_{N}+i\hat{q}\cdot\left(\frac{\vec{v}_{\mu}}{2}\times\vec{\sigma}_{L}\right)1_{N}\right]+\mathit{O}\left(\frac{q^{3}}{m_{N}^{3}}\right)$ & $2\frac{q}{m_{N}}\left(-\mathcal{O}_{1}-i\mathcal{O}_{2}^{f'}+\mathcal{O}_{3}^{f}\right)$   \\
31 & $\bar{\chi}_{e}\left(\gamma^{\mu}\cancel{\hat{q}}\gamma^{\nu}-\gamma^{\nu}\cancel{\hat{q}}\gamma^{\mu}\right)\chi_{\mu}\bar{N}\sigma_{\mu\nu}N$ & $-4i\left\{ \left(\hat{q}\cdot\vec{\sigma}_{L}\right)\left(\hat{q}\cdot\vec{\sigma}_{N}\right)-\vec{\sigma}_{L}\cdot\left[\hat{q}\times\left(\vec{v}_{N}\times\vec{\sigma}_{N}\right)\right]+\left(\frac{\vec{v}_{\mu}}{2}\cdot\vec{\sigma}_{L}\right)\left(\hat{q}\cdot\vec{\sigma}_{N}\right)\right\} +\mathit{O}\left(\frac{q^{2}}{m_{N}^{2}}\right)$ & $4\left(i\mathcal{O}_{6}+\mathcal{O}_{13}^{'}-\mathcal{O}_{14}^{f}\right)$   \\
32 & $\bar{\chi}_{e}\left(\gamma^{\mu}\cancel{\hat{q}}\gamma^{\nu}-\gamma^{\nu}\cancel{\hat{q}}\gamma^{\mu}\right)\chi_{\mu}\bar{N}\left(\gamma_{\mu}\frac{\cancel{q}}{m_{N}}\gamma_{\nu}-\gamma_{\nu}\frac{\cancel{q}}{m_{N}}\gamma_{\mu}\right)N$ & $8\frac{q}{m_{N}}\left\{ \left(\vec{\sigma}_{L}\cdot\vec{\sigma}_{N}\right)-\left(\hat{q}\cdot\vec{\sigma}_{L}\right)\left(\hat{q}\cdot\vec{\sigma}_{N}\right)\boldsymbol{-}\left(\hat{q}\cdot\vec{\sigma}_{L}\right)\left(\vec{v}_{N}\cdot\vec{\sigma}_{N}\right)\right.$ & $8\frac{q}{m_{N}}\left(\mathcal{O}_{4}+\mathcal{O}_{6}+i\mathcal{O}_{14}\right.$   \\
 &  & $\left.-i\left(\hat{q}\times\frac{\vec{v}_{\mu}}{2}\right)\cdot\vec{\sigma}_{N}+\left[\hat{q}\times\left(\frac{\vec{v}_{\mu}}{2}\times\vec{\sigma}_{L}\right)\right]\cdot\vec{\sigma}_{N}\right\} +\mathrm{\mathit{O}}\left(\frac{q^{3}}{m_{N}^{3}}\right)$ & $\left.-\mathcal{O}_{5}^{f}-i\mathcal{O}_{13}^{f'}\right)$\\
\end{tabular}
\end{ruledtabular}
  \end{minipage}
  }
  \end{table*}
\begin{table*}
\centering
\rotatebox{90}{
\begin{minipage}{\textheight}
\begin{ruledtabular}
\begin{tabular}{cccc}
$j$ & $\mathcal{L}_{\text{int}}^{j}$ & Pauli operator reduction & $\Sigma_{i}c_{i}\mathcal{O}_{i}$
\\
\hline 
33 & $\bar{\chi}_{e}\sigma^{\mu\nu}\chi_{\mu}\bar{N}\sigma_{\mu\nu}\gamma_{5}N$ & $-2\left(1_{L}\hat{q}\cdot\vec{\sigma}_{N}\right)-2i\vec{\sigma}_{L}\cdot\left(\hat{q}\times\vec{\sigma}_{N}\right)+\frac{q}{m_{N}}\left(\hat{q}\cdot\vec{\sigma}_{L}\right)1_{N}$ & $-2\mathcal{O}_{9}+2i\mathcal{O}_{10}-i\frac{q}{m_{N}}\mathcal{O}_{11}$   \\
 &  & $+2i\vec{\sigma}_{L}\cdot\left(\vec{v}_{N}\times\vec{\sigma}_{N}\right)+\vec{v}_{\mu}\cdot\vec{\sigma}_{N}+i\left(\vec{v}_{\mu}\times\vec{\sigma}_{L}\right)\cdot\vec{\sigma}_{N}+\mathit{O}\left(\frac{q^{2}}{m_{N}^{2}}\right)$ & $+2i\mathcal{O}_{12}+2\mathcal{O}_{8}^{f}+2i\mathcal{O}_{12}^{f}$   \\
34 & $\bar{\chi}_{e}\sigma^{\mu\nu}\chi_{\mu}\bar{N}\left(\frac{q_{\mu}}{m_{N}}\gamma_{\nu}-\frac{q_{\nu}}{m_{N}}\gamma_{\mu}\right)\gamma_{5}N$ & $\frac{q}{m_{N}}\left[2\vec{\sigma}_{L}\cdot\left(\hat{q}\times\vec{\sigma}_{N}\right)-2i1_{L}\left(\vec{v}_{N}\cdot\vec{\sigma}_{N}\right)\right.$ & $2\frac{q}{m_{N}}\left(-i\mathcal{O}_{9}-i\mathcal{O}_{7}\right.$   \\
 &  & $+i\left(\vec{v}_{\mu}\cdot\vec{\sigma}_{N}\right)-\left(\vec{v}_{\mu}\times\vec{\sigma}_{L}\right)\cdot\vec{\sigma}_{N}$ & $\left.+i\mathcal{O}_{8}^{f}-\mathcal{O}_{12}^{f}\right.$   \\
 &  & $\left.+\hat{q}\cdot\left(\vec{v}_{\mu}\times\vec{\sigma}_{L}\right)\left(\hat{q}\cdot\vec{\sigma}_{N}\right)-i\left(\hat{q}\cdot\vec{v}_{\mu}\right)\left(\hat{q}\cdot\vec{\sigma}_{N}\right)\right]+\mathit{O}\left(\frac{q^{3}}{m_{N}^{3}}\right)$ & $\left.-\mathcal{O}_{15}^{f}+i\mathcal{O}_{16}^{f'}\right)$   \\
35 & $\bar{\chi}_{e}\sigma^{\mu\nu}\chi_{\mu}\bar{N}\left(\frac{q_{\mu}}{m_{N}}v_{N\nu}-\frac{q_{\nu}}{m_{N}}v_{N\mu}\right)\gamma_{5}N$ & $-i\frac{q^{2}}{m_{N}^{2}}1_{L}\left(\hat{q}\cdot\vec{\sigma}_{N}\right)+\mathit{O}\left(\frac{q^{3}}{m_{N}^{3}}\right)$ & $-\frac{q^{2}}{m_{N}^{2}}\mathcal{O}_{10}$    \\
36 & $\bar{\chi}_{e}\sigma^{\mu\nu}\chi_{\mu}\bar{N}\left(\gamma_{\mu}\frac{\cancel{q}}{m_{N}}\gamma_{\nu}-\gamma_{\nu}\frac{\cancel{q}}{m_{N}}\gamma_{\mu}\right)\gamma_{5}N$ & $2\frac{q}{m_{N}}\left[-2i\hat{q}\cdot\vec{\sigma}_{L}1_{N}-2i\vec{\sigma}_{L}\cdot\vec{v}_{N}+2i\left(\hat{q}\cdot\vec{\sigma}_{L}\right)\left(\hat{q}\cdot\vec{v}_{N}\right)\right.$ & $-4\frac{q}{m_{N}}\left(\mathcal{O}_{11}+i\mathcal{O}_{8}+i\mathcal{O}_{16}^{'}\right.$   \\
 &  & $+\left.\frac{q}{m_{N}}\vec{\sigma}_{L}\cdot\left(\hat{q}\times\vec{\sigma}_{N}\right)+i\vec{v}_{\mu}\cdot\vec{\sigma}_{L}1_{N}\right]+\mathrm{\mathit{O}}\left(\frac{q^{3}}{m_{N}^{3}}\right)$ & $\left.+i\frac{q}{2m_{N}}\mathcal{O}_{9}-i\mathcal{O}_{7}^{f}\right)$   \\
37 & $\bar{\chi}_{e}\left(\hat{q}^{\mu}\gamma^{\nu}-\hat{q}^{\nu}\gamma^{\mu}\right)\chi_{\mu}\bar{N}\sigma_{\mu\nu}\gamma_{5}N$ & $i1_{L}\left(\hat{q}\cdot\vec{\sigma}_{N}\right)-2\vec{\sigma}_{L}\cdot\left(\vec{v}_{N}\times\vec{\sigma}_{N}\right)+2\left(\hat{q}\cdot\vec{\sigma}_{L}\right)\hat{q}\cdot\left(\vec{v}_{N}\times\vec{\sigma}_{N}\right)$ & $2\left(\frac{1}{2}\mathcal{O}_{10}-\mathcal{O}_{12}-\mathcal{O}_{15}\right.$   \\
 &  & $-\hat{q}\cdot\left(\vec{v}_{\mu}\times\vec{\sigma}_{L}\right)\left(\hat{q}\cdot\vec{\sigma}_{N}\right)+i\left(\hat{q}\cdot\vec{v}_{\mu}\right)\left(\hat{q}\cdot\vec{\sigma}_{N}\right)+\mathit{O}\left(\frac{q^{2}}{m_{N}^{2}}\right)$ & $\left.+\mathcal{O}_{15}^{f}-i\mathcal{O}_{16}^{f'}\right)$   \\
38 & $\bar{\chi}_{e}\left(\hat{q}^{\mu}\gamma^{\nu}-\hat{q}^{\nu}\gamma^{\mu}\right)\chi_{\mu}\bar{N}\left(\frac{q_{\mu}}{m_{N}}\gamma_{\nu}-\frac{q_{\nu}}{m_{N}}\gamma_{\mu}\right)\gamma_{5}N$ & $\frac{q}{m_{N}}\left[2i\vec{\sigma}_{L}\cdot\left(\hat{q}\times\vec{\sigma}_{N}\right)-1_{L}\left(\vec{v}_{N}\cdot\vec{\sigma}_{N}\right)\right.$ & $2\frac{q}{m_{N}}\left(\mathcal{O}_{9}-\frac{1}{2}\mathcal{O}_{7}\right.$   \\
 &  & $+\left(\vec{v}_{\mu}\cdot\vec{\sigma}_{N}\right)+i\left(\vec{v}_{\mu}\times\vec{\sigma}_{L}\right)\cdot\vec{\sigma}_{N}$ & $+\mathcal{O}_{8}^{f}+i\mathcal{O}_{12}^{f}$   \\
 &  & $\left.-i\hat{q}\cdot\left(\vec{v}_{\mu}\times\vec{\sigma}_{L}\right)\left(\hat{q}\cdot\vec{\sigma}_{N}\right)-\left(\hat{q}\cdot\vec{\sigma}_{N}\right)\left(\hat{q}\cdot\vec{v}_{\mu}\right)\right]+\mathit{O}\left(\frac{q^{3}}{m_{N}^{3}}\right)$ & $\left.+i\mathcal{O}_{15}^{f}+\mathcal{O}_{16}^{f'}\right)$   \\
39 & $\bar{\chi}_{e}\left(\hat{q}^{\mu}\gamma^{\nu}-\hat{q}^{\nu}\gamma^{\mu}\right)\chi_{\mu}\bar{N}\left(\frac{q_{\mu}}{m_{N}}v_{N\nu}-\frac{q_{\nu}}{m_{N}}v_{N\mu}\right)\gamma_{5}N$ & $-\frac{1}{2}\frac{q^{2}}{m_{N}^{2}}1_{L}\left(\hat{q}\cdot\vec{\sigma}_{N}\right)+\mathit{O}\left(\frac{q^{3}}{m_{N}^{3}}\right)$ & $\frac{i}{2}\frac{q^{2}}{m_{N}^{2}}\mathcal{O}_{10}$   \\
40 & $\bar{\chi}_{e}\left(\hat{q}^{\alpha}v_{\mu}^{\nu}-\hat{q}^{\nu}v_{\mu}^{\alpha}\right)\chi_{\mu}\bar{N}\sigma_{\alpha\nu}\gamma_{5}N$ & $2i1_{L}\hat{q}\cdot\vec{\sigma}_{N}+\hat{q}\cdot\left(\vec{v}_{\mu}\times\vec{\sigma}_{L}\right)\left(\hat{q}\cdot\vec{\sigma}_{N}\right)$ & $2\left(\mathcal{O}_{10}-\mathcal{O}_{15}^{f}\right.$   \\
 &  & $-i\left(\hat{q}\cdot\vec{v}_{\mu}\right)\left(\hat{q}\cdot\vec{\sigma}_{N}\right)+\mathit{O}\left(\frac{q^{2}}{m_{N}^{2}}\right)$ & $\left.+i\mathcal{O}_{16}^{f'}\right)$   \\
41 & $\bar{\chi}_{e}\left(\hat{q}^{\alpha}v_{\mu}^{\nu}-\hat{q}^{\nu}v_{\mu}^{\alpha}\right)\chi_{\mu}\bar{N}\left(\frac{q_{\alpha}}{m_{N}}\gamma_{\nu}-\frac{q_{\nu}}{m_{N}}\gamma_{\alpha}\right)\gamma_{5}N$ & $2\frac{q}{m_{N}}\left[-1_{L}\left(\vec{v}_{N}\cdot\vec{\sigma}_{N}\right)+\left(\vec{v}_{\mu}\cdot\vec{\sigma}_{N}\right)\right.$ & $2\frac{q}{m_{N}}\left(-\mathcal{O}_{7}+2\mathcal{O}_{8}^{f}\right.$   \\
 &  & $\left.-\left(\hat{q}\cdot\vec{v}_{\mu}\right)\left(\hat{q}\cdot\vec{\sigma}_{N}\right)\right]+\mathit{O}\left(\frac{q^{3}}{m_{N}^{3}}\right)$ & $\left.+2\mathcal{O}_{16}^{f'}\right)$   \\
42 & $\bar{\chi}_{e}\left(\hat{q}^{\alpha}v_{\mu}^{\nu}-\hat{q}^{\nu}v_{\mu}^{\alpha}\right)\chi_{\mu}\bar{N}\left(\frac{q_{\alpha}}{m_{N}}v_{N\nu}-\frac{q_{\nu}}{m_{N}}v_{N\alpha}\right)\gamma_{5}N$ & $\frac{q^{2}}{m_{N}^{2}}\left[-1_{L}\left(\hat{q}\cdot\vec{\sigma}_{N}\right)+i\hat{q}\cdot\left(\frac{\vec{v}_{\mu}}{2}\times\vec{\sigma}_{L}\right)\left(\hat{q}\cdot\vec{\sigma}_{N}\right)\right.$ & $\frac{q^{2}}{m_{N}^{2}}\left(i\mathcal{O}_{10}-i\mathcal{O}_{15}^{f}\right.$   \\
 &  & $\left.+\left(\hat{q}\cdot\frac{\vec{v}_{\mu}}{2}\right)\left(\hat{q}\cdot\vec{\sigma}_{N}\right)\right]+\mathit{O}\left(\frac{q^{4}}{m_{N}^{4}}\right)$ & $\left.-\mathcal{O}_{16}^{f'}\right)$   \\
43 & $\bar{\chi}_{e}\left(\gamma^{\mu}\cancel{\hat{q}}\gamma^{\nu}-\gamma^{\nu}\cancel{\hat{q}}\gamma^{\mu}\right)\chi_{\mu}\bar{N}\sigma_{\mu\nu}\gamma_{5}N$ & $-2\left[-2\vec{\sigma}_{L}\cdot\left(\hat{q}\times\vec{\sigma}_{N}\right)+i\frac{q}{m_{N}}\left(\hat{q}\cdot\vec{\sigma}_{L}\right)1_{N}\right.$ & $-4\left(i\mathcal{O}_{9}+\frac{q}{2m_{N}}\mathcal{O}_{11}\right.$   \\
 &  & $-2\left(\hat{q}\cdot\vec{\sigma}_{L}\right)\hat{q}\cdot\left(\vec{v}_{N}\times\vec{\sigma}_{N}\right)+i\vec{v}_{\mu}\cdot\vec{\sigma}_{N}-\left(\vec{v}_{\mu}\times\vec{\sigma}_{L}\right)\cdot\vec{\sigma}_{N}$ & $+\mathcal{O}_{15}+i\mathcal{O}_{8}^{f}-\mathcal{O}_{12}^{f}$   \\
 &  & $\left.+\hat{q}\cdot\left(\vec{v}_{\mu}\times\vec{\sigma}_{L}\right)\left(\hat{q}\cdot\vec{\sigma}_{N}\right)-i\left(\hat{q}\cdot\vec{v}_{\mu}\right)\left(\hat{q}\cdot\vec{\sigma}_{N}\right)\right]+\mathit{O}\left(\frac{q^{2}}{m_{N}^{2}}\right)$ & $\left.-\mathcal{O}_{15}^{f}+i\mathcal{O}_{16}^{f'}\right)$   \\
44 & $\bar{\chi}_{e}\left(\gamma^{\mu}\cancel{\hat{q}}\gamma^{\nu}-\gamma^{\nu}\cancel{\hat{q}}\gamma^{\mu}\right)\chi_{\mu}\bar{N}\left(\gamma_{\mu}\frac{\cancel{q}}{m_{N}}\gamma_{\nu}-\gamma_{\nu}\frac{\cancel{q}}{m_{N}}\gamma_{\mu}\right)\gamma_{5}N$ & $8\frac{q}{m_{N}}\left[-\left(\hat{q}\cdot\vec{\sigma}_{L}\right)1_{N}+\left(\vec{\sigma}_{L}\cdot\vec{v}_{N}\right)+i\frac{q}{2m_{N}}\vec{\sigma}_{L}\cdot\left(\hat{q}\times\vec{\sigma}_{N}\right)\right.$ & $8\frac{q}{m_{N}}\left(i\mathcal{O}_{11}+\mathcal{O}_{8}+\frac{q}{2m_{N}}\mathcal{O}_{9}\right.$   \\
 &  & $\left.-\left(\hat{q}\cdot\vec{\sigma}_{L}\right)\left(\hat{q}\cdot\vec{v}_{N}\right)-\left(\frac{\vec{v}_{\mu}}{2}\cdot\vec{\sigma}_{L}\right)1_{N}\right]+\mathrm{\mathit{O}}\left(\frac{q^{3}}{m_{N}^{3}}\right)$ & $\left.+\mathcal{O}_{16}^{'}-\mathcal{O}_{7}^{f}\right)$   \\
\end{tabular}
\end{ruledtabular}
  \end{minipage}
  }
  \end{table*}

In contrast to the WIMPs case, elastic $\mu \to e$ conversion occurs at a fixed momentum $q \sim m_{\mu}$. In addition, the electron 
is assumed to be fully relativistic.
Employing the operators described in \cite{Rule2023nuclear, haxton2023nuclear} with the appropriate order count and hierarchy scale $\frac{q}{m_L} > \frac{q}{m_N} \sim v_N > v_{\mu}$, we obtain the nucleon-level nonrelativistic reduction for elastic $\mu \to e$ conversion presented in Table \ref{tab: mu-to-e Nonrelativistic Reduction}, consisting with the $\mu \to e$ nonrelativistic operators:
\begin{equation}
\begin{split}
\mathcal{O}_{1} & \equiv 1_{L}1_{N} \text{,}\\
\mathcal{O}_{3} & \equiv 1_{L} i \hat{q}\cdot\left(\vec{v}_{N}\times\vec{\sigma}_{N}\right) \text{,}\\
\mathcal{O}_{4} & \equiv \vec{\sigma}_{L}\cdot\vec{\sigma}_{N} \text{,}\\
\mathcal{O}_{5} & \equiv \vec{\sigma}_{L}\cdot\left(i\hat{q}\times\vec{v}_N\right) \text{,}\\
\mathcal{O}_{6} & \equiv \left(i\hat{q}\cdot\vec{\sigma}_{L}\right)\left(i\hat{q}\cdot\vec{\sigma}_{N}\right) \text{,}\\
\mathcal{O}_{7} & \equiv 1_{L} \vec{v}_{N}\cdot\vec{\sigma}_{N} \text{,}\\
\mathcal{O}_{8} & \equiv \vec{\sigma}_{L} \cdot \vec{v}_{N} \text{,}\\
\mathcal{O}_{9} & \equiv \vec{\sigma}_{L} \cdot \left(i\hat{q}\times\vec{\sigma}_{N}\right) \text{,}\\
\mathcal{O}_{10} & \equiv 1_{L} i \hat{q}\cdot \vec{\sigma}_{N} \text{,}\\
\mathcal{O}_{11} & \equiv i \hat{q}\cdot \vec{\sigma}_{L} 1_{N} \text{,}\\
\mathcal{O}_{12} & \equiv \vec{\sigma}_{L}\cdot \left(\vec{v}_{N}\times\vec{\sigma}_{N}\right) \text{,}\\
\mathcal{O}_{13}^{'} & \equiv \vec{\sigma}_{L}\cdot \left[i\hat{q}\times\left(\vec{v}_{N}\times\vec{\sigma}_{N}\right)\right] \text{,}\\
\mathcal{O}_{14} & \equiv \left(i\hat{q}\cdot\vec{\sigma}_{L}\right)\left(\vec{v}_N\cdot\vec{\sigma}_{N}\right)\text{,}\\
\mathcal{O}_{15} & \equiv \left(i\hat{q}\cdot\vec{\sigma}_{L}\right) \left[i\hat{q}\cdot\left(\vec{v}_{N}\times\vec{\sigma}_{N}\right)\right]\text{,}\\%
\mathcal{O}_{16}^{'} & \equiv \left(i\hat{q}\cdot\vec{\sigma}_{L}\right) \left(i\hat{q}\cdot\vec{v}_{N}\right)\text{.}
\end{split}
\label{eq: NRET_operators_mu}
\end{equation}

Furthermore, accounting for the muon velocity $\vec{v}_{\mu}$ generates nuclear form factor corrections. These corrections are suppressed by the ratio of the average values of the lower and upper components of the muon, $\frac{\left<f\right>}{\left<g\right>}$, where $f$ and $g$ are the Coulomb Dirac solutions (see Ref.~ \cite{haxton2023nuclear} for details).  This leads to all the additional, smaller operators denoted by the superscript $f$:
\begin{equation}
\begin{split}
\mathcal{O}_{2}^{f'} & \equiv i\hat{q}\cdot\frac{\vec{v}_{\mu}}{2} 1_{N} \text{,}\\
\mathcal{O}_{3}^{f} & \equiv i \hat{q}\cdot\left(\frac{\vec{v}_{\mu}}{2}\times\vec{\sigma}_{L}\right) 1_{N} \text{,}\\
\mathcal{O}_{5}^{f} & \equiv \left(i\hat{q}\times\frac{\vec{v}_{\mu}}{2}\right)\cdot \vec{\sigma}_{N} \text{,}\\
\mathcal{O}_{7}^{f} & \equiv \frac{\vec{v}_{\mu}}{2}\cdot\vec{\sigma}_{L} 1_{N}\text{,}\\
\mathcal{O}_{8}^{f} & \equiv \frac{\vec{v}_{\mu}}{2}\cdot\vec{\sigma}_{N} \text{,}\\
\mathcal{O}_{12}^{f} & \equiv \left(\frac{\vec{v}_{\mu}}{2}\times\vec{\sigma}_{L}\right) \cdot\vec{\sigma}_{N} \text{,}\\
\mathcal{O}_{13}^{f'} & \equiv \left[i\hat{q}\times\left(\frac{\vec{v}_{\mu}}{2}\times\vec{\sigma}_{L}\right)\right] \cdot\vec{\sigma}_{N} \text{,}\\
\mathcal{O}_{14}^{f} & \equiv \left(\frac{\vec{v}_{\mu}}{2}\cdot\vec{\sigma}_{L}\right)\left(i\hat{q}\cdot\vec{\sigma}_{N}\right) \text{,}\\
\mathcal{O}_{15}^{f} & \equiv \left[i\hat{q}\cdot \left(\frac{\vec{v}_{\mu}}{2}\times\vec{\sigma}_{L}\right)\right]\left( i\hat{q}\cdot\vec{\sigma}_{N}\right) \text{,}\\
\mathcal{O}_{16}^{f'} & \equiv \left(i\hat{q}\cdot\frac{\vec{v}_{\mu}}{2}\right) \left(i\hat{q}\cdot\vec{\sigma}_{N}\right) \text{.}
\end{split}
\label{eq: NRET_operators_mu_f}
\end{equation}

In the leading order (dimension-six) of SMEFT, only the terms with $\bar{\chi}_{e}\sigma^{\mu\nu}\chi_{\mu}$ appear. These are the first four terms ($j \in \left\{21, 22, 23, 24\right\}$) listed in Table \ref{tab: mu-to-e Nonrelativistic Reduction}, and their $\gamma_5$ variations ($j \in \left\{33, 34, 35, 36\right\}$). These four terms, and separately, their $\gamma_5$ variations, should share the same leptonic coupling constant, to be multiplied by the appropriate tensor form factor, $g_T$ or $\tilde{g}_T^{\left(i\right)}$, depending on the specific nuclear Lorentz-invariant term from Eq. \eqref{eq: tensor_current}. All other $j$-terms require higher dimensions of SMEFT.

As highlighted in~\cite{Rule2023nuclear, haxton2023nuclear}, the tensor coupling introduces new operators absent in the non-tensor cases. Specifically, $\mathcal{O}_{3}$ and $\mathcal{O}_{13}^{'}$ manifest in the tensor-mediated interaction $\bar{\chi}_{e}\sigma^{\mu\nu}\chi_{\mu}\bar{N}\sigma_{\mu\nu}N$ (equivalent, up to its sign, to the interaction $\bar{\chi}_{e}i\sigma^{\mu\nu}{\gamma}^5\chi_{\mu}\bar{N}i\sigma_{\mu\nu}{\gamma}^5 N$ explored in~\cite{Rule2023nuclear, haxton2023nuclear}). This term possesses the potential to induce coherent effects, as discussed in the references above.
Here we find that $\mathcal{O}_{3}$ and $\mathcal{O}_{13}^{'}$ also appear separately: $\mathcal{O}_{3}$ in the terms $\bar{\chi}_{e}\left(\hat{q}^{\mu}\gamma^{\nu}-\hat{q}^{\nu}\gamma^{\mu}\right)\chi_{\mu}\bar{N}\sigma_{\mu\nu}N$ and $\bar{\chi}_{e}\left(\hat{q}^{\alpha}v_{\mu}^{\nu}-\hat{q}^{\nu}v_{\mu}^{\alpha}\right)\chi_{\mu}\bar{N}\sigma_{\alpha\nu}N$, and $\mathcal{O}_{13}^{'}$ in the term $\bar{\chi}_{e}\left(\gamma^{\mu}\cancel{\hat{q}}\gamma^{\nu}-\gamma^{\nu}\cancel{\hat{q}}\gamma^{\mu}\right)\chi_{\mu}\bar{N}\sigma_{\mu\nu}N$. As a result, their contributions can be distinguished.

In addition, we find that $\mathcal{O}_{12}$ and $\mathcal{O}_{15}$, which were not generated before, manifest in the tensor-mediated interaction $\bar{\chi}_{e}\left(\hat{q}^{\mu}\gamma^{\nu}-\hat{q}^{\nu}\gamma^{\mu}\right)\chi_{\mu}\bar{N}\sigma_{\mu\nu}\gamma_{5}N$, as well as separately: $\mathcal{O}_{12}$ in $\bar{\chi}_{e}\sigma^{\mu\nu}\chi_{\mu}\bar{N}\sigma_{\mu\nu}\gamma_{5} N$, and $\mathcal{O}_{15}$ in $\bar{\chi}_{e}\left(\gamma^{\mu}\cancel{\hat{q}}\gamma^{\nu}-\gamma^{\nu}\cancel{\hat{q}}\gamma^{\mu}\right)\chi_{\mu}\bar{N}\sigma_{\mu\nu}\gamma_{5}N$.
If one of these four new operators aligns with observed data, potentially enhanced by coherent effects, it would indicate that CLFV manifests a tensor nature, given that these operators are not linked to any other symmetry.

\paragraph*{Conclusions and outlook.}
NRET, as emphasized in~\cite{haxton2023nuclear}, offers a systematic framework for meticulous data analysis upon detecting CLFV or DM scattering, enabling the determination of the underlying nature of the new physics. Addressing the crucial gap in NRET, this study leverages an innovative technique to decompose antisymmetric tensor-type interactions, expanding NRET with vital tensor symmetries and creating a new avenue within the existing framework.

The inclusion of the missing tensor terms significantly impacts ongoing experiments.
A comprehensive understanding of NRET operators, including their involvement in the tensor symmetry, as outlined in Table~\ref{tab: WIMP Nonrelativistic Reduction} for DM and Table~\ref{tab: mu-to-e Nonrelativistic Reduction} for $\mu \to e$, is imperative for discerning tensor involvement in DM scattering or CLFV--a critical aspect overlooked by previous studies.

The author hopes that the provided comprehensive tables, featuring the previously missing tensor-mediator terms, and generating the last missing operators of the $\mu \to e$ conversion, will benefit the broad community exploring BSM physics.
These findings have the potential to contribute essential insights for ongoing and future experiments, deepening the understanding of tensor contributions in these new-physics processes.

Looking forward, while this letter discussed the tensor contact-level interaction between nuclear and BSM currents, it is important to also consider the potential impact of effective tensor couplings to photons and pions at the quark level (e.g., utilizing the electromagnetic field tensor $F^{\mu \nu}$), enabling contributions through particle exchange~\cite{dekens2019non, hoferichter2019nuclear}.
These new operators can also contribute via renormalization group evolution, potentially generating scalar interactions. Despite typically small coupling constants compared to contact interactions, their spin-independent nature allows significant contributions. Further exploration, including all possible tensor terms (Eq.~\eqref{eq: tensor_current}), is encouraged for a comprehensive understanding of their implications.

\paragraph*{Acknowledgments.}
The author would like to express gratitude towards the discussions held in the INT program "New physics searches at the precision frontier" (23-1b). The author also thanks Wick Haxton and Evan Rule for invaluable discussions and advice, and Vincenzo Cirigliano, Wouter Dekens, Doron Gazit, and Jerry Miller for their kind advice.
This work was supported in part by the U.S. Department of Energy (DOE) Office of Science, Office of Nuclear Physics, under award Number DE-FG02-97ER-41014, and the DOE Topical Collaboration “Nuclear Theory for New Physics (NTNP)”, award Number DE-SC0023663, as well as the Israel Academy of Sciences and Humanities, through the Post-Doctoral Fellowship in Nuclear Physics.
%
\bibliographystyle{unsrt}
\addcontentsline{toc}{section}{\refname}
\bibliography{main}

\end{document}